\documentclass[12pt]{article}
\usepackage{amsmath,amssymb}
\newcommand{\be}{\begin{equation}}
\newcommand{\ee}{\end{equation}}
\newcommand{\bea}{\begin{eqnarray}}
\newcommand{\eea}{\end{eqnarray}}

\def\P{Poincar\'e }

\begin{document}
\renewcommand {\theequation}{\thesection.\arabic{equation}}
\renewcommand {\thefootnote}{\fnsymbol{footnote}}
\vskip1cm
\begin{flushright}
\end{flushright}
\vskip1cm
\begin{center}
{\large\bf Twist and Spin-Statistics Relation\\ in Noncommutative
Quantum Field Theory} \vskip .7cm

{\bf{{Anca Tureanu}}

{\it High Energy Physics Division, Department of Physical Sciences,
University of Helsinki\\
\ \ {and}\\
\ \ Helsinki Institute of Physics,\\ P.O. Box 64, FIN-00014
Helsinki, Finland}}

\end{center}
\vskip1cm
\begin{abstract}

The twist-deformation of the Poincar\'e algebra as symmetry of the
field theories on noncommutative space-time with Heisenberg-like
commutation relation is discussed in connection to the relation
between a sound approach to the twist and the quantization in
noncommutative field theory. The recent claims of violation of
Pauli's spin-statistics relation and the absence of UV/IR mixing in
such theories are shown not to be founded.
\end{abstract}
\vskip1cm

\newpage
\section{Introduction}

Quantum field theories on noncommutative space-time with
Heisenberg-like commutation relation
\be\label{cr} [\hat x_\mu,\hat x_\nu]=i\theta_{\mu\nu}, \ee
where $\theta_{\mu\nu}$ is a antisymmetric constant matrix, have
been thoroughly investigated during the past years, after it had
been shown that they appear as low-energy limits of string theory in
a constant antisymmetric background field \cite{SW} (for a review,
see \cite{Szabo}). Two features of such noncommutative quantum field
theories (NC QFTs) render them specially interesting: the nonlocal
interaction and the violation of Lorentz invariance. The latter
feature is obvious by inspecting the commutation relation
(\ref{cr}), in which $\theta_{\mu\nu}$ does not transform under
Lorentz transformations. However, translational invariance is
preserved. The nonlocality becomes appearant when we write the
product of functions on the commutative counterpart of the
above-defined noncommutative space time, by the well-known
Weyl-Moyal correspondence, as a $\star$-product:
\be\label{star} \phi(x)\star
\psi(x)=\phi(x)e^{\frac{i}{2}\overleftarrow{\partial}_\mu\theta^{\mu\nu}\overrightarrow{\partial}_\nu}\psi(x)\
.\ee
The infinite number of derivatives involved in the $\star$-product
induces the nonlocality in the interaction terms of the Lagrangean.
One of the most interesting effects of the nonlocality is that the
high-energy (short-distance) behaviour is influenced by the topology
(long-distance effect) of the space-time \cite{CDP} and, as a
by-product, the UV/IR mixing \cite{UVIR}(an effect specific to
string theory) appears in the NC QFT on noncompact spaces.

Since NC QFTs are peculiar in many ways, it seemed natural to
investigate whether Pauli's spin-statistics relation is violated.
The idea was first mentioned in \cite{capri} and investigated in
more detail in \cite{CNT}, based on the equal-time commutation
relation of observables of free fields. The conclusion was that, at
least in NC QFTs with commutative time, i.e. $\theta_{0i}=0$, the
spin-statistics relation holds. The only hint to a possible
violation \cite{CNT} was in theories with light-like
noncommutativity, which are well-defined low-energy limits of string
theory, without unitarity problems \cite{unit}, but still acausal
due to the nonlocality in time \cite{caus}. Later on, in
\cite{CMNTV} the spin-statistics relation was shown to hold also in
the axiomatic formulation of NC QFT with commutative time
($\theta_{0i}=0$). In all these investigations, and indeed in the
whole literature on the subject, it was assumed that the fields are
still in the representations of the Lorentz group, though it was
obvious the Lorentz symmetry was violated. The justification of this
treatment came when it was realized that NC QFT has
twisted-Poincar\'e symmetry and as such it has exactly the same
representation content as the usual Poincar\'e-invariant QFT
\cite{CKT} (since the twist does not affect the algebra of the
generators of the \P symmetry, but only their action in a tensor
product of representations, i.e. their co-product, see also
\cite{Wess}.)

Recently, however, the twisted \P symmetry of NC QFT was exploited
in a new manner, leading to claims that the spin-statistics relation
does not hold in NC QFT \cite{Bala1} and the UV/IR mixing disappears
\cite{Bala2}. In short, it was stated that in order to be "{\it
compatible} with the deformed action of the \P group" \cite{Bala2},
the standard commutation relations of creation and annihilation
operators have to be also deformed, and not identical with those of
the usual QFT, as it is taken in the "traditional NC
QFT"\footnote{Though NC QFT is a relatively new field of research,
we shall call the well-known approach based on the Weyl-Moyal
correspondence "traditional NC QFT", to differentiate it from the
"deformed-statistics" approach of \cite{Bala1,Bala2}.}.

In this letter we shall prove that the truly twisted-Poincar\'e
compatible approach to NC QFT is the traditional one, in which the
spin-statistics relation holds and UV/IR mixing remains, while the
new approach \cite{Bala1,Bala2} is in effect a commutative theory.

\section{Twisted \P symmetry and canonical\\ quantization}
We shall not repeat the construction of the twisted \P algebra, but
refer the reader to \cite{CKT, CPrT} and references therein. It
suffices here to say that an Abelian twist element is introduced
\be\label{twist}{\cal F}=e^{\frac{i}{2}\theta^{\mu\nu}P_\mu\otimes
P_\nu}\ ,\ee
where $P_\mu$ is the generator of the (Abelian) translation
subalgebra of the \P algebra and $P_\mu\otimes P_\nu$ is the tensor
product of generators. This twist element does not affect the
algebra of the \P generators, but deforms the action of the Lorentz
generators $M_{\mu\nu}$ in the tensor product of representations.
Moreover, the twist changes the multiplication in the algebra of
representations (in the case of field theory, the algebra of
representations is the algebra of the fields). ${\cal A}$ is the
algebra of functions of the coordinates of the Minkowski space,
which carries the following representation of the \P algebra:
\bea [P_\mu, \phi(x)]&=&({\cal P}_\mu\ \phi)(x)=i\partial_\mu
\phi(x)\ ,\cr [M_{\mu\nu},\phi(x)]&=&({\cal M}_{\mu\nu}\
\phi)(x)=i(x_\mu\partial_\nu-x_\nu\partial_\mu)\phi(x)\ .\eea
The product of the elements of the algebra ${\cal A}$ (the product
of fields) is deformed upon twisting as
\bea\label{twist_product} \phi(x)\star
\psi(y)&=&m\left(e^{-\frac{i}{2}\theta^{\mu\nu}P_\mu\otimes
P_\nu}\phi(x)\otimes \psi(x)\right)\cr
&=&e^{-\frac{i}{2}\theta^{\mu\nu}{\cal P}^x_\mu{\cal
P}^y_\nu}\phi(x)\psi(y)=e^{\frac{i}{2}\theta^{\mu\nu}\partial^x_\mu
\partial^y_\nu}\phi(x)\psi(y)\ , \eea
which is the generalization of the $\star$-product (\ref{star})
defined through Weyl-Moyal correspondence (see also \cite{Szabo}).
By taking in (\ref{twist_product}) $\phi(x)=x_\mu$ and
$\psi(x)=x_\nu$, one obtains immediately the Moyal bracket of
coordinates, $[x_\nu,x_\nu]_\star=i\theta_{\mu\nu}$. The
$\star$-product of two functions written as Fourier expansions
\be\label{fourier} \phi(x)=\int d^4p\ \tilde \phi(p)e^{-ipx}\ ,\ \ \
\psi(y)=\int d^4k\ \tilde \psi(k)e^{-iky} \ee
is the one induced also by the Weyl-Moyal correspondence
\bea\label{WM_product} \phi(x)\star \psi(y)&=&\int d^4p\ d^4k\
\tilde \phi(p)\tilde \psi(k)e^{-ipx}\star e^{-iky}\cr &=&\int d^4p\
d^4k\ \tilde \phi(p)\tilde
\psi(k)e^{-\frac{i}{2}p^\mu\theta_{\mu\nu}k^\nu}e^{-i(px+ky)}\ .\eea
where $\tilde \psi(k)\tilde \phi(p)=\tilde \phi(p)\tilde \psi(k)
$.

The Lagrangean of a noncommutative field theory, e.g., the NC
$\lambda\Phi^4$ theory, is built therefore with $\star$-products
instead of usual products of field:
\be{\cal
L}_\star(x)=\frac{1}{2}\partial^\mu\Phi(x)\star\partial_\mu\Phi(x)-\frac{1}{2}m^2\Phi(x)\star\Phi(x)
-\frac{\lambda}{4!}\Phi(x)\star\Phi(x)\star\Phi(x)\star\Phi(x)\ ,\ee
and it is twisted-\P invariant.

Thus by the deformation of \P algebra with the twist (\ref{twist}),
one reproduces the construction of NC field theory by Weyl-Moyal
correspondence. It is well-known that under the integration over the
whole space-time one $\star$-product drops out, therefore the action
of the free NC field is the same as the action of the corresponding
commutative field. They also satisfy the same equation of motion,
therefore one can use the same mode-expansion for the free NC
hermitian scalar field as for the commutative one:
\be\label{mode_exp}\Phi(x)=\int\frac{d^3p}{(2\pi)^{3/2}2E_p}\left[c({\bf
p})e^{-ipx}+c^\dagger({\bf p})e^{ipx}\right]\ ,\ \ \
p_0=E_p=\sqrt{{\bf p}^2+m^2}\ .\ee

\subsection{Traditional approach in the light of twist}

{\it Quantization in operator formulation versus path-integral
formulation}

When one attempts the canonical quantization of a free NC field, one
has to be careful with the canonical commutation relation imposed on
the field $\Phi(x)$ and its conjugated momentum
$\Pi(x)=\frac{\partial {\cal L}}{\partial\dot\Phi(x)}$, since such a
commutation relation involves {\it products} and thus the
multiplication, to be compatible with the twisted \P symmetry of the
NC space-time, has to be effected by the $\star$-product. The
$\star$-product induces an infinite nonlocality in the noncommmuting
directions, i.e. an infinite speed of propagation of signal in these
directions, leading to an alteration of the causality condition, for
example, which should be formulated as non-correlation of the events
out of each other's light-wedge \cite{LAG} and not light-cone.
Therefore, the equal-time (canonical) commutation relation of the
commutative case
\be\label{cr_com}\left[\Phi(t,{\bf x}),\Pi(t,{\bf
y})\right]=i\delta({\bf x}-{\bf y})\ee
has to be also suitably modified to take into account the
nonlocality in the NC directions.

However, one can bypass many difficulties of the operator
formulation and quantize the theory using the path integral
approach, subsequently drawing also conclusions on the canonical
quantization procedure.

The straightforward generalization of the generating functional
$W(J)$ to the NC case is:
\bea\label{W(J} W(J)&=&\int \prod_{x}{\cal
D}\mu(u_1(x),...,u_n(x))\cr&\times&\exp\left[i\int d^4x\left({\cal
L}_{\star}(x)+u_1(x)\star J_1(x)+...+u_n(x)\star
J_n(x)\right)\right]\cr&=&\int \prod_{x}{\cal
D}\mu(u_1(x),...,u_n(x))\cr&\times&\exp\left[i\int d^4x\left({\cal
L}_0(x)+{\cal
L}_{I\star}(x)+u_1(x)J_1(x)+...+u_n(x)J_n(x)\right)\right]\ , \eea
where $u_i(x)$, $i=1,...,n$ are the fields entering the
noncommutative Lagrangean density ${\cal L}_\star$, which is
obtained from its commutative counterpart by replacing the usual
products by $\star$-products, as indicated in the previous section.
The fact that the $\star$-product drops out in the quadratic terms
under the integration over the whole space-time is taken into
account. The expression (\ref{W(J}) is twisted \P covariant, since
the integration measure does not change under twist.

%
%

It is common knowledge in the traditional approach to NC QFT that by
using the generating functional (\ref{W(J}) one obtains an S-matrix
expansion equivalent to the Dyson expansion obtained in the operator
formulation by using noncommutative form of the interaction
Lagrangean, but the usual commutation relations of the creation and
annihilation operators in the in-fields, e.g.
\bea\label{cr_ann} \left[c({\bf p}), c^\dagger({\bf
q})\right]=2E_p\delta({\bf p}-{\bf q})\,\cr
 \left[c({\bf p}), c({\bf q})\right]=\left[c^\dagger({\bf p}), c^\dagger({\bf q})\right]=0\ ,\eea
for the scalar field (\ref{mode_exp}).

In this quantization framework for the noncommutative fields, the
requirement of commutative time, $\theta_{0i}=0$, is natural, since
this is the situation which does not violate unitarity \cite{unit},
nor causality \cite{caus}. In this case, one can always choose a
frame of reference (the inertial frames of reference are not
equivalent, since Lorentz invariance is violated) in which only two
space directions are noncommutative, e.g.
$\theta_{12}=-\theta_{21}\neq 0$, all the other components of the
matrix $\theta_{ij}$ being zero. In this configuration, which we
shall adopt throughout, noncommutativity is in the plane
$(x_1,x_2)$, while the coordinates $x_0$ and $x_3$ commute among
themselves and with the others.

With the hint that the commutation relation of the creation and
annihilation operators remains the same as in the commutative case
(\ref{cr_ann}), we turn to the twisted-\P analog of (\ref{cr_com}),
to see which alterations the noncommutativity induce on it. The
simple equal-time $\star$-commutator of the free field
(\ref{mode_exp}) with its canonically conjugated momentum
\be\label{ETCR}\left[\Phi(t,{\bf x}),\Pi(t,{\bf y})\right]_\star\
,\ee
turns out to be operator valued on the Hilbert space of quantum
states. However, the matrix elements of (\ref{ETCR}) between two
states of equal total momentum $|\Psi_i(P_\mu)\rangle$ and
$|\Psi_f(P_\mu)\rangle$ (i.e., the matrix elements corresponding to
physical transitions) have the familiar form
\be\label{etcr_nc}\langle\Psi_f(P_\mu)|\left[\Phi(t,{\bf
x}),\Pi(t,{\bf y})\right]_\star|\Psi_i(P_\mu)\rangle=i\delta({\bf
x}-{\bf y})\ \ee
if $|\Psi_i(P_\mu)\rangle=|\Psi_f(P_\mu)\rangle$ (diagonal matrix
elements). If $|\Psi_i(P_\mu)\rangle\neq|\Psi_f(P_\mu)\rangle$, then
the r.h.s. of (\ref{etcr_nc}) vanishes, just as in the commutative
case.

The $\star$-commutator of free scalar fields at two space-time
points is also operator-valued on the Hilbert space of states, but
its diagonal matrix elements are
\be\langle\Psi_f(P_\mu)|\left[\Phi(x),\Phi(y)\right]_\star|\Psi_i(P_\mu)\rangle=i\Delta_c(x-y),\
\ \ \ |\Psi_i(P_\mu)\rangle=|\Psi_f(P_\mu)\rangle \ee
where $\Delta_c(x-y)$ is the causal function:
\be\label{causal_delta}\Delta_c(x-y)=-\frac{i}{2(2\pi)^3}\int
d^4k\epsilon(k_0)\delta(k^2-m^2)e^{-ikx}\ , \ee
implying
\be\label{causal_free}\langle\Psi_f(P_\mu)|\left[\Phi(x),\Phi(y)\right]_\star|\Psi_i(P_\mu)\rangle=0\
,\ \ \ \mbox{for}\ \ (x_0-y_0)^2-({\bf x}-{\bf y})^2<0\ ,\ \ \ \
|\Psi_i(P_\mu)\rangle=|\Psi_f(P_\mu)\rangle\ . \ee
Again, for $|\Psi_i(P_\mu)\rangle\neq|\Psi_f(P_\mu)\rangle$, the
r.h.s. of (\ref{causal_free}) vanishes, as expected by comparison
with the commutative case.

Thus, for the NC free scalar hermitian field, the physically
meaningful matrix elements of the $\star$-equal time commutation
relation (\ref{etcr_nc}) or $\star$-commutator of fields
(\ref{causal_free}) have the same expressions as in the commutative
case. As a consequence, when deriving the Feynman rules for the NC
case, one obtains the same propagator as in the corresponding
commutative theory.

The situation becomes different when considering NC interactions. A
similar matrix element has been calculated for the Heisenberg field
of an interacting scalar field theory, in one loop, but the result
was generalized to any order in perturbation theory \cite{Chu}. The
only difference was that the commutator was calculated without
$\star$-product between the fields, but, as it will be shown later,
for Heisenberg fields it is not important whether one takes their
commutator with $\star$-product (as one rigorously should) or
without. Thus, according to \cite{Chu} and having in view the above
comment, for scalar Heisenberg fields one obtains the diagonal
matrix elements
\be\langle\Psi(P_\mu)|\left[\Phi_H(x),\Phi_H(y)\right]_\star|\Psi(P_\mu)\rangle=
i\Delta(x_0-y_0,x_3-y_3)F(x_1-y_1,x_2-y_2,\theta_{12}), \ee
where $F(x_1-y_1,x_2-y_2,\theta_{12})$ is a function which may
vanish in a finite number of points and
$$
\Delta_c(x_0-y_0,x_3-y_3)=-\frac{i}{2(2\pi)^3}\int
dk^0dk^3\epsilon(k_0)\delta(k_0^2-k_3^2-m^2)e^{-i(k^0x_0+k^3x_3)}
$$
is the analog of the causal function $\Delta_c(x-y)$, implying the
"light-wedge" causality condition (see \cite{LAG})
\be\label{light-wedge}\langle\Psi(P_\mu)|\left[\Phi_H(x),\Phi_H(y)\right]_\star|\Psi(P_\mu)\rangle=0\
,\ \ \ \mbox{for}\ \ (x_0-y_0)^2-(x_3-y_3)^2<0. \ee
Correspondingly, the diagonal matrix elements of the
$\star$-commutator of a scalar Heisenberg field with its canonically
conjugated momentum shall read
\be\label{etcr_H}\langle\Psi(P_\mu)|\left[\Phi_H(t,{\bf
x}),\Pi_H(t,{\bf y})\right]_\star|\Psi(P_\mu)\rangle=i\delta(x_3-
y_3)G(x_1-y_1,x_2-y_2,\theta_{12})\ ,\ee
where $G(x_1-y_1,x_2-y_2,\theta_{12})$ has similar properties with
$F(x_1-y_1,x_2-y_2,\theta_{12})$. Again, for
$|\Psi_i(P_\mu)\rangle\neq|\Psi_f(P_\mu)\rangle$, the r.h.s. of both
(\ref{light-wedge}) and (\ref{etcr_H}) are zero.

It appears thus that the physical matrix elements of commutators of
{\it nonlocal} operators, like (\ref{light-wedge}), vanish outside
the light-wedge, as expected from nonlocality considerations.

In brief, this is the essence of the traditional approach to NC QFT,
leading to the usual spin-statistics relation \cite{CNT,CMNTV} and
UV/IR mixing \cite{UVIR}.

\subsection{Deformed-statistics approach}

In \cite{Bala1,Bala2}, it was argued that, since the Fourier
transform $\tilde \phi(p)$ of a function $\phi(x)$ is also a linear
representation of the momentum generator $P_\mu$ of the \P algebra,
the product of Fourier transforms should be also deformed upon
twisting the \P algebra. Pursuing this line of thought, it was
claimed that the creation and annihilation operators satisfy
deformed commutation relations. We do not repeat the argumentation
here, but merely cite their outcome, using the common convention for
Fourier expansion (\ref{fourier})\footnote{The conventions used in
\cite{Bala1,Bala2} are different from the conventions usually used
in the literature and in this letter, but amounting, essentially, to
taking the Fourier expansion as $\phi(x)=\int d^4p \tilde
\phi(p)e^{ipx}$ and $\theta_{\mu\nu}\rightarrow-\theta_{\mu\nu}$ in
(\ref{cr}).}.

A free quantum scalar field of mass $m$ is expanded as:
\be\label{bala_modes} \Phi(x)=\int d\mu({\bf p})\left[a({\bf
p})e^{-ipx}+a^\dagger({\bf p})e^{ipx}\right]\ ,\ \ \ee
where $d\mu({\bf p})=\frac{d^3p}{(2\pi)^{3/2}2E_p}$ and
$p_0=E_p=\sqrt{{\bf p}^2+m^2}$. The deformed creation and
annihilation operators $a^\dagger({\bf p})$ and $a({\bf p})$ were
represented in \cite{Bala1} in terms of the nondeformed ones,
$c^\dagger({\bf p})$ and $c({\bf p})$, as
\bea\label{a_op} a({\bf p})=c({\bf
p})e^{-\frac{i}{2}p_\mu\theta^{\mu\nu}P_\nu}\cr a^\dagger({\bf p})=
e^{\frac{i}{2}p_\mu\theta^{\mu\nu}P_\nu}c^\dagger ({\bf p})\ ,\eea
where $c^\dagger({\bf p})$ and $c({\bf p})$ satisfy the usual
commutation relations (\ref{cr_ann}) and
\be P_\mu=\int d\mu({\bf p})p_\mu c^\dagger ({\bf p}) c({\bf
p})=\int d\mu({\bf p})p_\mu a^\dagger ({\bf p}) a({\bf p})\ee
is the quantum momentum operator, generating a linear representation
on the creation and annihilation operators:
\bea\label{op-repres} \left[P_\mu,a({\bf p})\right]=-p_\mu\ a({\bf
p}),\ \ \ \ \left[P_\mu,c({\bf p})\right]=-p_\mu\ c({\bf p})\cr
\left[P_\mu,a^\dagger({\bf p})\right]=p_\mu\ a({\bf p}),\ \ \ \
\left[P_\mu,c^\dagger({\bf p})\right]=p_\mu\ c({\bf p})\ . \eea
Using (\ref{a_op}) and (\ref{op-repres}), we can write down the
deformed commutation relations of the creation and annihilation
operators $a^\dagger({\bf p})$ and $a({\bf p})$:
\bea\label{deformed_cr} a({\bf p})a({\bf q})&=&e^{-iq\theta p}a({\bf
q})a({\bf p})\ ,\cr
 a^\dagger({\bf p})a^\dagger({\bf q})&=&e^{-iq\theta p}a^\dagger({\bf q})a^\dagger({\bf p})\ ,\cr
 a({\bf p})a^\dagger({\bf q})&=&e^{iq\theta p}a^\dagger({\bf q})a({\bf p})+2E_p\delta({\bf p}-{\bf q})\ ,\cr
a^\dagger({\bf p}) a({\bf q})&=&e^{iq\theta p}a({\bf
q})a^\dagger({\bf p})-2E_p\delta({\bf p}-{\bf q})\ ,\eea
where the notation $q\theta p=q_\mu\theta^{\mu\nu}p_\nu$ is used. A
typical term of the product of fields $\Phi\star\Phi$ is
\be\label{bala-prod} a({\bf p})a({\bf q})e^{-ipx}\star e^{-iqx}\ .
\ee

It is obvious that the multiparticle-states
$|n\rangle=a^\dagger({\bf p}_1)a^\dagger({\bf p}_2)\cdots
a^\dagger({\bf p}_n)|0\rangle$ described by the scalar field with
the above quantization are not symmetric, therefore they do not
satisfy the Bose-Einstein statistics. Therefore, it was argued
\cite{Bala1} that the spin-statistics relation, in this case, does
not hold. Moreover, with the commutation rules of the creation and
annihilation operators (\ref{deformed_cr}) it was concluded in
\cite{Bala2} that the S-matrix in this case turns out to be
identical to the corresponding one in the commutative case,
consequently the UV/IR mixing does not appear.

At this stage it is interesting to see what is the equal-time
commutation relation of fields and conjugated momenta and what is
the causality condition analog to (\ref{light-wedge}) to which these
new commutation rules for creation and annihilation operators
(\ref{deformed_cr}) lead. A straightforward calculation gives:
\be\label{bala-etcr}  \left[\Phi(t,{\bf x}),\Pi(t,{\bf
y})\right]_\star=i\delta({\bf x}-{\bf y}) \ee
and
\be\left[\Phi(x),\Phi(y)\right]_\star=i\Delta_c(x-y)\ ,\ee
where $\Delta_c(x-y)$ is the usual four-dimensional causal function
(\ref{causal_delta}), leading to
\be\label{light-cone}\left[\Phi(x),\Phi(y)\right]_\star=0\ ,\ \ \
\mbox{for}\ \ (x_0-y_0)^2-(\vec x-\vec y)^2<0. \ee

It is obvious from the calculations that the $\star$-product of the
exponentials from the mode expansion (\ref{bala_modes}) is exactly
canceled by the deformed commutation rules of the creation and
annihilation operators (\ref{deformed_cr}). In effect, for any two
functions $\phi(x)=\int d^4p\ \tilde \phi(p)e^{-ipx}$ and
$\psi(y)=\int d^4q\ \tilde \psi(q)e^{-iqy}$, with $\tilde
\phi(p)\tilde \psi(q)=e^{-iq\theta p}\tilde \psi(q)\tilde \phi(p)$,
according to \cite{Bala1}, one has:
\bea \phi(x)\star \psi(y)&=&\int d^4p\ d^4q\ \tilde \phi(p)\tilde
\psi(q)e^{-ipx}\star e^{-iqy}=\int d^4p\ d^4q\ \tilde \phi(p)\tilde
\psi(q)e^{-\frac{i}{2}p\theta q}e^{-ipx-iqy}\cr &=&\int d^4p\ d^4q\
e^{-iq\theta p}\tilde \psi(q)\tilde \phi(p)e^{-\frac{i}{2}p\theta
q}e^{-ipx-iqy}\cr &=&\int d^4p\ d^4q\ \tilde \psi(q)\tilde
\phi(p)e^{-\frac{i}{2}q\theta p}e^{-iqy-ipx}=\int d^4p\ d^4q\ \tilde
\psi(q)\tilde \phi(p)e^{-iqy}\star e^{-ipx}\cr&=& \psi(y)\star
\phi(x)\ .\eea
Thus the $\star$-product of two functions of $x$ is commutative,
thereby rendering any $\star$-product of quantum fields in the
S-matrix expansion of an interacting theory precisely as in the
corresponding commutative theory. With the same argument, the
commutator of Heisenberg fields of a theory with interactions will
be the same as in the corresponding commutative case:
\be\label{H_light-cone}\left[\Phi_H(x),\Phi_H(y)\right]_\star=0\ ,\
\ \ \mbox{for}\ \ (x_0-y_0)^2-({\bf x}-{\bf y})^2<0. \ee

Now the reason for the absence of the UV/IR mixing is cleared up:
the theory constructed in \cite{Bala1,Bala2} is a {\it local} one,
in spite of the nonlocal $\star$-product specific to noncommutative
field theories.

\vskip 0.5cm

{\it Spin-statistics relation}

As for the spin-statistics relation, the situation is slightly more
subtle. Indeed, the deformation of the commutation relations
(\ref{cr_ann}) into the form (\ref{deformed_cr}) implies a deformed
statistics for a scalar field, which might be interpreted as a
violation of the spin-statistics relation. However, since the theory
under consideration in \cite{Bala1} is a {\it local} relativistic
scalar field theory, as is clearly seen from (\ref{H_light-cone}),
it does fulfill all the requirements of Pauli's {\it spin-statistics
theorem} \cite{Pauli}, but it {\it contradicts} its conclusion, the
spin-statistics relation. This is an obvious indication that the
corresponding NC theory is not properly quantized and raises the
suspicion whether the deformed statistics (\ref{deformed_cr}) as a
quantization procedure is introduced in a manner compatible with the
twist, investigated in the next section.

\subsection{Twisted multiplication of representations of $P_\mu$}

To require that the Fourier transforms, as linear representations of
the momentum generator $P_\mu$ undergo the action of the twist in a
product just like any other representations of $P_\mu$, e.g. the
exponentials $e^{-ipx}$, is certainly legitimate. However, in this
case one should require that the product of Fourier transforms and
exponentials corresponding to {\it different} momenta be also
deformed. Therefore, the {\it usual} product between elements of
different algebras of representation of $P_\mu$, e.g.
\be \tilde \phi(p)e^{-iqx}=e^{-iqx}\tilde \phi(p)\ ,\ee
as used in \cite{Bala1,Bala2}, is not compatible with the concept of
twist as a general abstract operation.

In effect, any tensor product of representations should be affected
by the twist. In the above case, according to the general rule
(2.10) of \cite{CKT}:
\bea \tilde \phi(p)\star e^{-iqx}=m_t(\tilde \phi(p)\otimes
e^{-iqx})&=&m(e^{-\frac{i}{2}\theta^{\mu\nu}P_\mu\otimes
P_\nu}\tilde \phi(p)\otimes
e^{-iqx})\\&=&m(e^{-\frac{i}{2}\theta^{\mu\nu}{\cal
P}^p_\mu\otimes{\cal P}^x_\nu}\tilde \phi(p)\otimes
e^{-iqx})=e^{-\frac{i}{2}p_\mu\theta^{\mu\nu}q_\nu}\tilde
\phi(p)e^{-iqx}\nonumber,\eea
since
\be [P_\mu,\tilde \phi(p)]={\cal P}^p_\mu\tilde \phi(p)=p_\mu\
\tilde \phi(p)\ ,\ \ \ \ [P_\nu,e^{-iqx}]={\cal P}^x_\nu
e^{-iqx}=q_\nu\ e^{-iqx}\ .\ee
Obviously, with this rule, the Fourier transform of a function is
preserved as usual, since
$$
\tilde \phi(p)\star e^{-ipx}=\tilde \phi(p) e^{-ipx}\ .
$$
Then the product of two functions defined on the Minkowski space,
written as Fourier expansions (\ref{fourier}), will be indeed given
by the action of the twist in the product of four representations
$$
\tilde \phi(p)\star e^{-ipx}\star \tilde \psi(q)\star
e^{-iqy}=m_t(\tilde \phi(p)\otimes e^{-ipx}\otimes \tilde
\psi(q)\otimes e^{-iqy}).
$$
The only nontrivial product under the twist is the one in the
middle, $e^{ipx}\star \tilde \psi(q)$. Consequently
\bea \phi(x)\star \psi(y)&=&\int d^4p\ d^4q\ \tilde \phi(p)\star
e^{-ipx}\star \tilde \psi(q)\star e^{-iqy}= \int d^4p\ d^4q\ \tilde
\phi(p)\left(e^{-ipx}\star \tilde \psi(q)\right) e^{-iqy}\cr&=&\int
d^4p\ d^4q\ \tilde \phi(p) \tilde
\psi(q)e^{-\frac{i}{2}p_\mu\theta^{\mu\nu}q_\nu}e^{-ipx-iqy}\cr&\equiv&\int
d^4p\ d^4q\ \tilde \phi(p) \tilde \psi(q)e^{-ipx}\star e^{-iqy}\ ,
\eea
i.e., by considering the action of the twist on {\it all}
multiplications of the representations of the momentum generator
$P_\mu$, one obtains exactly the $\star$-product induced by the
Weyl-Moyal correspondence (\ref{WM_product}). In other words, the
traditional approach to NC QFT described in Subsection (2.1), with
usual spin-statistics relation and the UV/IR mixing, is the one
which is truly compatible with the twisted \P symmetry.

\section{Conclusions}

In this letter we show that by consistently handling the twisted \P
algebra as the symmetry of NC field theory, namely the deformed
multiplication in the tensor product of representations of the
momentum generator $P_\mu$, one is lead to a quantization procedure
in operator formulation which preserves Pauli's spin-statistics
relation, for theories with commutative time. Moreover, the
traditional approach based on the Weyl-Moyal correspondence for NC
models is entirely recovered, together with its UV/IR mixing
problems.

The claims of spin-statistics violation in theories with twisted \P
symmetry are thereby naturally rejected. By this approach one
obtains essentially {\it commutative}, local quantum field theories
with deformed statistics, thus contradicting Pauli's spin-statistics
theorem. The drawback of the construction leading to such claims is
that the quantization of the NC field theory is performed by simply
imposing (deformed) commutation relations between the creation and
annihilation operators, instead of following a canonical
quantization procedure. Technically, in a twist-deformed product of
quantum fields $\Phi\star\Phi$, the {\it deformed commutation
relations} of creation and annihilation operators, e.g. $a({\bf
p})a({\bf q})=e^{-iq\theta p}a({\bf q})a({\bf p})$, is not
equivalent to the {\it deformed product} of those operators, $a({\bf
p})\star a({\bf q})=e^{-\frac{i}{2}p\theta q}a({\bf p})a({\bf q})$.
Thus, at present, the only possible case in which violation of
spin-statistics relation might appear within NC QFT seems to be in
theories with light-like noncommutativity,
$\theta^{\mu\nu}\theta_{\mu\nu}=0$ \cite{CNT}. The latter has no
problems with unitarity and can be obtained as a low-energy limit of
string theory.

NC QFT has an infinite range of nonlocality of interaction in the
noncommuting directions, i.e. the nonlocality is not restricted to a
finite range, such as the Planck scale. It is certainly desirable to
find means of restricting this nonlocality, while still preserving
some of its peculiarities which mimic stringy effects, in a more
manageable context. However, the deformation of statistics (i.e.
deformation of the commutation relations between creation and
annihilation operators) in NC quantum field theories with
commutative time is not the way to attain this scope.

\vskip 1.5cm {\bf{Acknowledgements}}

We are grateful to M. Chaichian, C. Montonen, K. Nishijima and P.
Pre\v{s}najder for several illuminating discussions.

\end{document}